\documentclass[11pt]{article}
\usepackage{amsfonts,latexsym}
\newcommand{\cleqn}{\setcounter{equation}{0}}
\newcommand{\clth}{\setcounter{theorem}{0}}
\newcommand {\sectionnew}[1]{\section{#1}\cleqn\clth}
\newcommand{\beq}{\begin{equation}}
\newcommand{\eeq}{\end{equation}}

\newcommand{\beqa}{\begin{eqnarray}}
\newcommand{\eeqa}{\end{eqnarray}}
\newcommand{\beaa}{\begin{eqnarray*}}
\newcommand{\ben}{\begin{eqnarray*}}
\newcommand{\eaa}{\end{eqnarray*}}
\newcommand{\een}{\end{eqnarray*}}

\newcommand{\text}{\textrm}
\newcommand \nc {\newcommand}
\nc \proof {\noindent {\em{Proof.\/ }}}
\nc \qed {$\Box$\hfill}
\newtheorem{theorem}{Theorem}[section]
\newtheorem{lemma}[theorem]{Lemma}
\newtheorem{proposition}[theorem]{Proposition}
\newtheorem{corollary}[theorem]{Corollary}
\newtheorem{definition}[theorem]{Definition}
\newtheorem{example}[theorem]{Example}
\newtheorem{remark}[theorem]{Remark}
\newtheorem{conjecture}[theorem]{Conjecture}
\newtheorem{question}[theorem]{Question}
\nc \bth[1] { \begin{theorem}\label{t#1} }
\nc \ble[1] { \begin{lemma}\label{l#1} }
\nc \bpr[1] { \begin{proposition}\label{p#1} }
\nc \bco[1] { \begin{corollary}\label{c#1} }
\nc \bde[1] { \begin{definition}\label{d#1}\rm }
\nc \bex[1] { \begin{example}\label{e#1}\rm }
\nc \bre[1] { \begin{remark}\label{r#1}\rm }
\nc \bcon[1] { \begin{conjecture}\label{con#1}\rm }
\nc \bque[1] { \begin{question}\label{que#1}\rm }
\nc {\eth} { \end{theorem} }
\nc {\ele} { \end{lemma} }
\nc {\epr} { \end{proposition} }
\nc {\eco} { \end{corollary} }
\nc {\ede} { \end{definition} }
\nc {\eex} { \end{example} }
\nc {\ere} { \end{remark} }
\nc {\econ} { \end{conjecture} }
\nc {\eque} { \end{question} }
\nc \eqref[1] {{\rm{(\ref{#1})}}}
\nc \thref[1]{Theorem \ref{t#1}}
\nc \leref[1]{Lemma \ref{l#1}}
\nc \prref[1]{Proposition \ref{p#1}}
\nc \coref[1]{Corollary \ref{c#1}}
\nc \deref[1]{Definition \ref{d#1}}
\nc \exref[1]{Example \ref{e#1}}
\nc \reref[1]{Remark \ref{r#1}}
\nc \conref[1]{Conjecture \ref{con#1}}

\def \a {\alpha}
\def \b {\beta}

\def \Rset {{\mathbb R}}
\def \Cset {{\mathbb C}}

\def \ad { {\mathrm{ad}} }

\def \p { {\partial}}

\nc \Wr {Wr}
\nc \GRN { \Gr^{(N)} }
\nc \GRA[1] { \Gr_A^{(#1)} }   
\nc \GRAN { \GRA{N} }
\nc \GrA[1] { \Gr_A(#1) }\nc \GrAa { \GrA{\alpha} }
\nc \GRB[1] { \Gr_B^{(#1)} }   
\nc \GRBN { \GRB{N} }
\nc \GrB[1] { \Gr_B(#1) }
\nc \GrBb { \GrB{\beta} }
\nc \GRMB[1] { \Gr_{MB}^{(#1)} }   
\nc \GRMBN { \GRMB{N} }
\nc \GrMB[1] { \Gr_{MB}(#1) }
\nc \GrMBb { \GrMB{\beta} }

\begin{document}

\title{{\LARGE\bf{ Strictly nilpotent elements and bispectral operators
 in the Weyl algebra}}}
\author{
E. ~Horozov
\thanks{E-mail: horozov@fmi.uni-sofia.bg. Supported
by Grant MM 1003-2000 of NFSR of Bulgaria}
\\ \hfill\\ \normalsize \textit{Department of Mathematics and Informatics,}\\
\normalsize \textit{ Sofia University, 5 J. Bourchier Blvd., Sofia
1126, Bulgaria }       }
 \maketitle
\date{}

\begin{abstract}
In this paper we give another characterization of the strictly
nilpotent elements in the Weyl algebra, which (apart from the
polynomials) turn out to be all bispectral operators with
polynomial coefficients. This also allows to reformulate in terms
of bispectral operators the famous conjecture, that all the
endomorphisms of the Weyl algebra are automorphisms (Dixmier,
Kirillov, etc).

\end{abstract}


AMS classification numbers: primary 47E05; secondary 16S32.
    \newpage
\setcounter{section}{-1}
\sectionnew{Introduction}

 In a recent series of papers \cite{H1, H2, HM} there has been
  made an attempt to
broaden the classification of bispectral operators that was
started by J.J.Duistermaat and F.A.Gr\"unbaum in \cite{DG} and
continued by G.Wlson in \cite{W1}. The present paper could be
considered as yet another step in this direction. I believe,
however, that the results could be of interest also to specialists
in other areas of research, in particular connected to the Weyl
algebra $A_1$.
 For this reason I will try to present the material
without using any specific knowledge on bispectral operators. To
explain the main message of the paper let me first recall some
definitions and results.

In what follows we will consider the algebra $A_1$ in its standard
realization - i.e. as the algebra of differential operators with
polynomial coefficients.
  An element $M \in A_1$ is said to {\em act nilpotently} on a
non-constant element $H\in A_1$, when there exists a positive
integer $m$ such that $\ad_M^m(H)=0$. An element is {\em strictly
nilpotent} if it acts nilpotently on all elements of $A_1$.
Slightly paraphrasing Dixmier  the strictly nilpotent elements are
characterized  as {\em {those that belong to the orbits of the
operators with constant coefficients under the action of the
automorphism group $Aut(A_1)$ of $A_1$}}.

Now I pass to the other main object of the present paper - the
bispectral operators. They have been introduced by F.A.Gr\"unbaum
(cf. \cite{G1}) in his studies on applications of spectral
analysis to medical imaging. Later it turned out that they are
connected to several actively developing areas of mathematics and
physics - the KP-hierarchy, infinite-dimensional Lie algebras and
their representations, particle systems, automorphisms of algebras
of differential operators, non-commutative geometry, etc. (see
e.g. \cite{BHY1, BHY2, BHY3, BW, BW1, BW2, K, W1, W2, MZ}, as well
as the papers in the proceedings volume of the conference in
Montr\'eal \cite{BP}).

 An ordinary differential operator
$L(x,\p_x)$ is called bispectral if it has an eigenfunction
$\psi(x,z)$, depending also on the spectral parameter $z$, which
is at the same time an eigenfunction of another differential
operator $\Lambda(z,\partial_z)$ now in the spectral parameter
$z$.  In other words we look for operators $L$ , $\Lambda$ and a
function $\psi(x,z)$ satisfying equations of the form:

 \beqa
   &&L\psi=f(z)\psi, \label{1.1} \hfill\\
    &&\Lambda\psi=\theta(x)\psi, \label{1.2} \hfill
    \eeqa
where the functions are defined in some open sets of $\Cset $ and
$ \Cset^2$. A simple consequence of the above definition is that
the bispectral operator $L$ {\em acts nilpotently} on the function
$\theta(x)$. This is the well known $\ad$-condition from \cite{DG}
which is widely believed to be not only necessary but also a
sufficient condition for the bispectrality of $L$, provided that
$L$ is normed as follows:

\beq
  L=\p^N+ \sum_{j=0}^{N-2} V_j\p^j, \label{1.3}
   \eeq
   i.e. $V_N=1$ and $V_{N-1}=0$. In fact for operators in $A_1$
   the first condition, i.e. $V_N=1$ suffices as by conjugating
   $L$ by $exp(Q(x))$ with an appropriate polynomial $Q(x)$ the second
   condition can be achieved remaining in $A_1$.
      In what follows we will also use this relaxed norming condition, i.e. we
   will assume that $L$ has the form:

\beq
  L=\p^N+ \sum_{j=0}^{N-1} V_j\p^j, \label{1.3'}
   \eeq
 We are ready to formulate our main results.

\bth{1.2}
   A differential operator from $A_1$  of the form \eqref{1.3'} is bispectral if
   and only if it is strictly nilpotent.
   \eth

This result will be an easy consequence of the following theorem
which will be formulated purely in terms of the Weyl algebra, i.e.
without referring to bispectral operators.

\bth{1.1}
         An element $L\in A_1$ is strictly
         nilpotent if and only if it is either a polynomial in $x$ or
         has the form \eqref{1.3'} and acts nilpotently on some
         non-constant polynomial $\theta(x)$.
     \eth
In other words the strictly nilpotent operators (of non-zero
order) are exactly those that satisfy the $\ad$-condition and
\eqref{1.3'}.

The theorems announced above constitute the main body of the
paper.

Next I will explain some further connections between the Weyl
algebra and bispectral operators.

In the fundamental paper \cite{DG} Duistermaat and Gr\"unbaum
classified the bispectral operators of order two. Roughly speaking
they are Darboux transformations of some Bessel operators the only
exception being the Airy operator. Another important result, due
to G.Wilson \cite{W1}, is that all bispectral operators of rank 1
are Darboux transformations of operators with constant
coefficients. (This is not the original form of Wilson's theorem
but a well known reformulation, cf. e.g. \cite{H1}). In
\cite{BHY3} we suggested a general scheme for producing bispectral
operators by application of Darboux transformations out of
"simple" ones which apparently works for all differential
operators (see also \cite{H1}). Now it seems that the most
difficult problem in the classification of bispectral operators is
not to perform Darboux transformations but to find a reasonable
class of operators that could be considered "simple". The results
of the present paper show that in $A_1$ all "simple" operators are
those that satisfy the {\em canonical commutation relation} (CCR
for brevity)

\beq
  [L,P]=1; \label{1.3a}
       \eeq


 One can reinterpret the main results also as follows.

\bpr{1.1} The centralizer $C(L)$ of any bispectral operator in
$A_1$ is generated by an element $L^{'}$, which together with some
other element $P$ satisfy the CCR \eqref{1.3a}.
  \epr
   In view of this re-interpretation it is tempting to conjecture the opposite:

  \bcon {1.1}
  If two elements $L_0$ and $P$ satisfy the CCR \eqref{1.3a} they
  are bispectral.
  \econ
  The results of the present paper allow to easily show that the
  above conjecture is equivalent to the famous conjecture of
  Dixmier-Kirillov :

       \bcon{1.2} If  two operators $L$ and $P$ satisfy CCR then they
generate $A_1$. In other words every endomorphism of the Weyl
algebra is automorphism.
   \econ

This equivalence is demonstrated at the end of the last section.

To make the presentation less dependent on \cite{Dx} we recall
some of the results that are needed in section 1. Then in the next
section we give the proofs of the above results.

\sectionnew{Preliminaries on the Weyl algebra}

Before proceeding with the main results we briefly recollect some
of the notions and results from \cite{Dx}. The first Weyl algebra
$A_1$ is an associative algebra generated over a field $F$ by two
elements $p$ and $q$, subject to the canonical commutation
relation (CCR) $[p,q]=1$ . In this paper $F$ will be always
$\Cset$, although most of the results can be reformulated for more
general fields. In most of the paper $A_1$ will be realized as the
algebra of differential operators of one variable $x$ with
polynomial coefficients, where $p=\p$ and $q=x$. A major tool in
the the study of $A_1$ is the introduction of suitable
filtrations. Let $G=\sum a_{i,j}q^ip^j$ and let $E(G)$ be the set
of all pairs $(i,j)$, such that $a_{i,j}\neq0$. If $\rho$ and
$\sigma$ are two real numbers put

$$v_{\rho,\sigma}(G)= \sup_{(i,j)\in E(G)}(\rho i +\sigma j).$$
    Denote by $E(G,\rho,\sigma)$ the set of pairs $(i,j)$, such
    that $\rho i + \sigma j= v_{\rho,\sigma}(G)$. With each $G$, $\rho$ and
$\sigma$ we associate a polynomial $f(X,Y)$ (in the commuting
variables $X$ and $Y$) as follows:

\beq
  f(X,Y)= \sum_{a_{i,j}\in E(G,\rho,\sigma)}a{i,j}X^iY^j. \label{2.1}
   \eeq
$f$ will be called {\em the polynomial $(\rho,\sigma)$-associated
with $G$}. Now we are ready to recall {\bf Lemma.7.3} from
\cite{Dx} on the "normal form" of a polynomial $f$ associated with
an element $G$. To avoid unnecessary for us terminology we recall
it for the particular case that we need. Namely we consider that
$G\in A_1$ act on  $M\in A_1$ nilpotently. Let $\rho,\sigma$ be
positive integers and let $f$ and $g$ be the polynomials
$(\rho,\sigma)$-associated with $G$ and $M$ correspondingly and
put $v=v_{\rho,\sigma}(G)$ and $w=v_{\rho,\sigma}(M)$.

\ble{2.1}
   Assume that $v+w > \rho +\sigma$ and that $f$ is not a
   monomial. Then one of the following cases holds:

   (i) $f^w$ is proportional to $g^v$;

   (ii) $\sigma>\rho$, $\rho$ divides $\sigma$ and
   \beq
          f(X,Y)= \lambda X^{\a}(X^{{\sigma}/{\rho}}+\mu Y)^{\b};
          \label{2.2}
          \eeq

   (iii) $\rho>\sigma$, $\sigma$ divides $\rho$ and
   \beq
          f(X,Y)= \lambda Y^{\a}(Y^{{\rho}/{\sigma}}+\mu X)^{\b}
          \label{2.3};\eeq

          (iv) $\sigma=\rho$,  and
   \beq
          f(X,Y)= \lambda(\mu X+\nu Y)^{\a}(\mu^{'}X+\nu^{'}
          Y)^{\b},          \label{2.4}
           \eeq
           where $\lambda,\mu, \mu{'}, \nu, \nu{'}\in Cset$ and $\a$, $\b$ are
         non-negative integers.
 \ele

 At the end we introduce (after Dixmier) the following notations.
 Let $S(\p)$ be a polynomial in $\p$. Then the  automorphism
 $\Phi_S$, given by:

 \beq
          \Phi_S= e^{\ad_{S(\p)}} \label{2.5}
     \eeq
is well defined. In the same way for a polynomial $R(x)$ one
defines:
  \beq
 \Psi_R= e^{\ad_{R(x)}}.\label{2.6}
   \eeq
      A fundamental result from \cite{Dx} is the following theorem, which will be
      used in the present paper.

      \bth{2.1}
      The group $Aut(A_1)$ of automorphisms  of $A_1$
      is generated by the automorphisms \eqref{2.5} and \eqref{2.6}.
      \eth

      \sectionnew{Proofs}
In this section we are going to give the proofs of the results
from the Introduction. No doubt the central one is \thref{1.1}
from which the rest are easy consequences.

 The proof of \thref{1.1}
 uses an induction reminiscent of "Fermat's method of infinite
descent", the main step of which in our case  reduces the number
of the factors of the order $N$ of $L$. This will be done as
follows. We can suppose that $L$ depends on $x$, i.e. it is not a
polynomial in $\p$.
 First choose appropriately $\rho$ and $\sigma$ in such a way
that the polynomial $f$, $(\rho, \sigma)$-associated with $L$ has
two terms, the first one being $Y^N$. Then show that $f$ has the
form from \leref{2.1}, {\em (iii)} with $\a=0$. The next step is
to apply an appropriate automorphism of $A_1$, sending our
operator $L$ to another one with similar properties but reducing
the number of factors of the order $N$.


 We need some preparations for the proof.
  Write $L$ in the form:
     \beq
   L=\sum_{(i,j) \in E(L)} a_{i,j}x^i\p^j \label{3.2}
   \eeq
  with $a_{0,N}=1$, $a_{i,N}=0$, $i>0$.
   We would like to consider  now the non-trivial cases when
  at least one point $(i,j)$ with
  $i>0$ belongs to $E(L)$, i.e. we assume that $L$ depends
  non-trivially on $x$. Assuming that we  will explain how to
  choose the  weights $\rho$ and $\sigma$ to fit our purposes.
   Draw the line in the plane $\Rset^2$ passing
   through the point (0,N)
  and at least one other  point, say $(k,m)$ with $k>0$
   and such that all other points remain below or on the line.
    Then one can choose $\rho$ and
  $\sigma$ to be one non-zero  solution in integers of the
  equation  $N \sigma= k \rho + m \sigma$. The solution does not depend
  on the specific $(k,m)$. This gives that the
  polynomial  $f$, $(\rho, \sigma)$-associated with $L$ has the
  form:

  \beq
         f(X,Y)= Y^N + a_{k,m}X^k Y^m + \ldots, \quad a_{k,m}
         \neq0.         \label{3.3}
         \eeq
 Here we have chosen the pair $(k,m)$ so that $k$ is the greatest possible.
  Our main concern will be to study the polynomial $f$ associated with
  $L$.
   Introduce also the following object.
Let $M$ be an element from the orbit of $x$, which does not
commute with $L$ and has the form:

    \beq
           M=\Psi_{R_1}\circ \Phi_{S_1}\circ \ldots \circ \Psi_{R_l}\circ
           \Phi_{S_l}(x),  \label{3.1}
    \eeq
where $R_j$, $S_j$ are polynomials with $degR_j\geq3$,
$degS_j\geq3$. The number $l$ could be zero. In this case $M=x$.

   \ble{3.1} Assume that $L$ is given as in \eqref{1.3} and that
   it acts nilpotently on a non-constant polynomial $\theta(M)$
  in $M$, where $M$ is given in \eqref{3.1}.
     Let $k>1$ in the above expression \eqref{3.3} of $f$.
     Then $v_{\rho,\sigma}(L)> \rho+ \sigma$.
   \ele
  \proof Writing $f$ in the form:

  \beq
      f= Y^m(Y^{N-m}+a_{k,m} X^k + \ldots), \quad a_{k,m} \neq 0
      \label{3.4}
     \eeq
     we can choose $\sigma = k$ and $\rho = N-m$. If $m=0$ and
     $k=N=2$ then according to \cite{Dx}, {\bf Lemma.7.4} the
     element $L$ is strictly semisimple and hence acts nilpotently
      only on elements of its centralizer $C(L)$, which
      cannot be true since $M$ does not commute with $L$.
      (A simple independent proof is also possible,
      cf. e.g. \cite{H2}. Hence we
    can assume that either $m>0$ or $\max{(N,k)}\geq 3 $. Then we
    have

    \ben
        v_{\rho,\sigma}(L)= N \sigma = Nk \geq N+ k \geq (N-m) +
        k= \rho + \sigma.
    \een
 In the case of $m>0$ the second inequality in the above chain is
 strict, while in the case of $\max{(N,k)}\geq 3 $ the first inequality
 is strict (recall that both  $N\geq 2$ and $k\geq 2$).\qed

 Next find a normal form for the polynomial associated with $L$.

 \ble{3.2} Assume that $L$ has at least one
 nonconstant coefficient $ V_j(x)$ and
 satisfies the conditions of \leref{3.1}.
 Then there exist numbers $\rho$ and $\sigma$, such that
 the
 polynomial $(\rho, \sigma)$-associated with $L$ has the form
 \beq
   f=(Y^r-\lambda X)^k, \quad \lambda\neq 0 \label{3.5}
     \eeq
    \ele
  \proof We choose the integers $\rho$ and $\sigma$ as explained
  above so that the $f$ has the form \eqref{3.4}. This is possible
  due to the assumption that $L$ has at least one nonzero coefficient.
   First assume that
  in \eqref{3.1} we have $k>1$. In this case according to
  \leref{3.1} $v_{\rho,\sigma}(L)> \rho+ \sigma$. Hence
   we can
  apply \leref{2.1}. Note that the polynomial
  $(\rho,\sigma)$-associated with $\theta(M)$ has the form $g=\gamma X^l$,
  hence the case {\em(i)} is ruled out.
  If we assume that the case {\em(ii)} of \leref{2.1}
  holds than expanding $f$ we see that the second coefficient
  in the expansion of   $L$ in $\p$, i.e. the coefficient
  $a_{N-1}$ is not zero which contradicts \eqref{1.3}. By the same
  reason the case {\em (iv)} is possible only with $\lambda= \nu =
  \nu^{'}=1$,
   $\mu = - \mu^{'}\neq 0$ and $\a=\b$ or equivalent to it. Then
   applying the a linear automorphism $\Psi$, defined by $\Psi(\p) =
   \p + \mu x$, $\Psi(x) = x $ we can bring the polynomial $f$
   into the form $f= Y^{\a}(Y + 2\mu X)^{\a}$, keeping $g$
   untouched (but not $\theta(M)$). Finally,  if $f$ has the form
   \eqref{3.1} with $k=1$ then it is exactly
   $f= Y^m(Y^{N-m} +\lambda X)$. Summing up the above two cases
   as well as    {\em (ii)} we get that in general $f$ has the
   form:

   \beq
      f=Y^n(Y^r-\lambda X)^k, \quad k\geq 1, \quad \lambda \neq 0
      \label{3.7}
   \eeq
  Now we want to show that $n=0$. Perform the automorphism $\Phi=\Phi_{S_0}$
where $S_0(\p)= \lambda^{-1} \frac{\p^{r+1}}{r+1}$. This
automorphism maps $L$ into a new element $\Phi(L)$ with a new
polynomial $f_0$ $(\rho, \sigma)$-associated with $L$ of the form:

     \beq
          f_0 = (- \lambda)^k X^k Y^n, \label{3.8}
     \eeq
   while the polynomial associated with $\Phi(\theta(M))$ will become
   \beq
         g_0 = \gamma_0( X + \lambda^{-1}Y^r)^l, \quad
         \gamma_0 \neq 0. \label{3.9}
   \eeq
    We are going to use that $\Phi(L)$ acts nilpotently on
   $\Phi(\theta(M))$. In what follows we will drop the non-essential
    coefficients $\gamma^{'}$ and $(-\lambda)^k$.  Let us compute
    consecutively $\ad_{\Phi(L)}^s(\Phi(\theta(M)))$  with
    $s=1,\ldots$. As we will be interested only on the terms with highest
    weight we will drop the rest. Then we have

    \beqa
          &&\Phi(L)= \p^n x^k + \ldots \label{3.10} \hfill\\
          &&\Phi(\theta(M))= (x+\lambda^{-1} \p^r)^k+ \ldots.
           \label{3.11}\hfill
    \eeqa
  Expand the highest weight terms \eqref{3.11} as

  \beq
  (x+\lambda^{-1}\p^r)^l = \sum c_j^l x^j\lambda^{-l+j} \p^{r(l-j)} +
  \ldots,\label{3.12}
  \eeq
 By linearity we have
    \beq
       \ad_{\Phi(L)}^s (\Phi(\theta(M)))=
   ad_{\p^n x^k}^s (x+\lambda^{-1}\p^r)^l =
    \sum_{j=0}^l c_j^l \lambda^{-l+j}ad_{(\p^n x^k)}^s
( x^j \p^{r(l-j)})+
   \ldots  \label{3.13}
       \eeq
     We will  consider separately two cases: with $k\geq n$ and
       $n\geq k$.

     1)  Let $k\geq n$. Simple computation gives that

     $$
 ad_{\p^n x^k}^s (x^l)= \prod_{j=0}^{s-1}[nl + (k-n)j]\p^{s(n-1)}
 x^{l+ s(k-1)} + \ldots
 $$
   Having in mind that $n\geq 1$ and $k-n \geq 0$ we get that the
   coefficient at the highest power of $x$ is always positive for
   any $s\geq 1$, which shows that \eqref{3.12} cannot be zero.
   This contradicts the fact that $L$ acts nilpotently on
   $\theta(M)$.

   2) Let  $n\geq k$. We have

   $$ ad_{\p^n x^k}^s (\p^{lr})= (-1)^s \prod_{j=0}^{s-1}[lrk +
(n-k)j] \p^{s(n-1)+lr}x^{s(k-1)} + \ldots $$ By the same argument
the coefficient at the
 highest power in $\p$ is not zero for any
$s$. This shows that either $n=0$ or $k=0$. But from the
assumption \eqref{3.1} it follows that $k$ cannot be zero.\qed

Now we perform the main induction step.

    \ble{3.3} Assume the conditions of the above lemma. Then
     there exists a polynomial $S(\p)$ with $degS(\p)=r+1\geq 3$,
   such that the image $L_1$ of $L$ under the action of the
   corresponding automorphism $\Phi_R$ has the form:

   \beq
       L_1=  \Phi_S(L) = (-\lambda)^k x^k + \sum_{j<k}c_j(\p)x^j, \quad
   c_j(\p) \in \Cset[\p], \quad
         c_{k-1}\equiv 0. \label{3.6}
   \eeq
 \ele
     \proof Use the obvious fact that the elements $\p$ and $\p^r - \lambda
     x$ are generators of $A_1$. Then
     \leref{3.2} shows that the element $L$ can be written
     in the form:

     \beq
   L= (\p^r - \lambda x)^k +
   \sum_{j=0}^{k-1}b_{i,j}(\p^r - \lambda x)^j \p^i \label{3.14'}
     \eeq
     Apply the automorphism $\Phi$ from the proof of the
     previous lemma, i.e. $\Phi(\p)=\p$,
     $\Phi(x)= x+ \lambda^{-1}\p^r$. Put $L_1=\Phi(L)$,
     $\theta_1=\Phi(\theta(M))$. Then one can write $L_1$
     (dropping the non-essential constant factor) in the form:

     \beq
     L_1= x^k + \sum_{j=0}^{k-1} b_j(\p)x^j,\quad
    b_j(\p) \in \Cset[\p] . \label{3.14}
     \eeq
     Notice that in the above expression all the terms after $x^k$
     have weights less than $N$ in the chosen filtration. In particular
     for $b_{k-1} \neq 0$ we have:
  \beq
     v_{\rho,\sigma}(b_{k-1}(\p)x^{k-1})< N=kr. \label{3.15}
  \eeq
      Assume that $b_{k-1}\neq 0$.
      Having in
     mind that our filtration can be chosen so that $\rho=r$, $\sigma=1$
     the  inequality \eqref{3.15} can be rewritten as $deg b_{k-1}+r(k-1)< kr$.
     This shows that the degree of $b_{k-1}$ is less than $r$. By an
     appropriate automorphism $\Phi_0 $ we can kill $b_{k-1}$.
     The composition $\Phi_0\circ \Phi$ is the automorphism
     $\Phi_S$ we are looking for. Notice that  $S(\p)= c \p^{r+1} +
     S_0(\p)$,
      where $deg S_0 \leq r$ and $c \neq 0$. Hence the degree of $S$ is
     exactly $r+1$. \qed
\vskip5pt
    Let us give the proofs of the main results.

     \proof of \thref{1.1}.
\vskip5pt
     If the
  second coefficient $V_{N-1}$ of $L$ is not zero then apply
  appropriate automorphism $\Psi_R$, where $R^{'}=-V_{N-1}$. This
  will bring our operator $L$ into the situation of \leref{3.1}
  with $M=x$.
     If the number
    $k$ from \eqref{3.5} is equal to 1 then  $L$
    is the generalized Airy operator, hence in the orbit of $\p$.
    So assume  that the number $k>1$. If we
    assume that all the coefficients of $L$ are constant then the
    theorem is again proven. Now assume that at least one coefficient
    of $L$ is not constant. Then according to \leref{3.3} we can
    find an automorphism $\Phi_S$ which sends $L$ into
    \eqref{3.6}. Notice that the operator $L_1$ from \eqref{3.6}
    has the properties of $L$ required by \leref{3.2}
    (with $x$ and $-\p$ exchanging their
    places) but its order $k$ is strictly less than the order $N$
    of $L$. This shows that after a finite number of steps we will
    come to either a polynomial in $x$ or in $\p$, thus proving
    that $L$ is in their orbits. \qed

  The next corollary follows from the proof of the last theorem
  (but not from  the theorem as stated).
 \bco{3.1}
Let the operator $L$ satisfy the conditions of \thref{1.2}. Then
it has the form similar to \eqref{3.3}. More precisely $L$ is a
polynomial in an element $K$ in the form
  \beq
  K=  \Phi_{1}\circ \Psi_{1}\circ \ldots \circ \Psi_{l}\circ
           \Phi_{l+1}(x), \label{form1}
  \eeq
  where $\Psi_j= \Psi_{R_j}$, $\Phi_j= \Phi_{S_j}$,
  $j=1,\ldots, l$ and the polynomials have degrees $\geq 3$. The
  automorphism $\Phi_{l+1}$ is either of the same form or is defined
  by $\Phi_{l+1}(x)=\p$, $\Phi_{l+1}(\p)= -x$.
  \eco

\vskip5pt
     Let us give the \proof of \thref{1.2}.
     \vskip5pt
   If $L$ is
  normalized as in \eqref{1.3} and bispectral it acts nilpotently
  on some nonconstant polynomial $\theta(x)$. Hence by \thref{1.1}
  it is strictly nilpotent. The opposite also follows easily.
  Suppose that $L$ belongs to the orbit of some nonconstant
  polynomial in $\p$, say $Q(\p)$. We need to consider only the
  case when $L$ is not a polynomial in $x$. Then there exists an
  automorphism $\phi$ such that $L=Q(\phi(\p))$. Denote by
  $L_0$ the operator $\phi(\p)$.

   Let $b_0$ be the
  standard anti-involution:

  \beq
  b_0(x) = \p_z,\quad \quad b_0(\p_x)= z. \label{3.16}
  \eeq
  (As usually treating bispectral operators we use different variables -
  $x$ and $z$ for the two copies of $A_1$.) Now define (cf.
  \cite{BHY3}) the anti-involution $b=b_0\circ \phi$. It is enough
 to show that $L_0$ is bispectral. Then the bispectrality of $L$
 will follow immediately as $L$ is a polynomial of $L_0$.
   We have

   \beq
   L_0=b^{-1}(z) = \phi\circ b_0^{-1}(z) = \phi(\p_x). \label{3.17}
   \eeq
   Define

   \beq
       \Lambda = b(x) = b_0 \circ \phi^{-1}(x). \label{3.18}
          \eeq
     We have only to exhibit the wave function
    $\psi(x,z)$, so that \eqref{1.1} and \eqref{1.2} are satisfied
    with $L_0$, $\Lambda$,
    $f(z)=z$ and $\theta(x)=x$ . We can always assume that $L_0$
    is normalized as in \eqref{1.3}. Otherwise we can apply appropriate
    automorphism as explained above and bring it to this form. The point is
    that we would like to use \coref{3.1}, which assures that the
    the polynomials, defining the automorphism $\phi$ are of
    degree $3$ or more except for $\Phi_{l+1}$. Then we can apply
     the theorem from
    \cite{BHY4} which gives the wave function in explicit form.
   \qed


 In what follows it would
be convenient to consider the polynomials of $x$ also bispectral.
(In fact allowing the wave function to be distribution they are,
cf. \cite{BHY3}.

In view of \thref{1.2} it is obvious that the centralizer of each
bispectral operator $L$ is generated by $\phi(\p)$, where $\phi$
is the automorphism, defining $L$ from the proof of \coref{3.1}.
Introduce also the operator $\phi(x)$. Then obviously they satisfy
the CCR

   \beq
     [\phi(\p),\phi(x)]=1 \label{4.1}
    \eeq
    This gives the \proof of \prref{1.1}.\qed

It is tempting to try to prove the opposite, i.e. \conref{1.1}.
This conjecture seems to be difficult to prove. The results of the
present paper allow to show that it is equivalent to \conref{1.2}.

  We  will give the simple proof of the equivalence of the two
   conjectures in the following form .

       \bpr{4.1} Let $L$, $P$ be two operators from $A_1$ that satisfy
       the CCR \eqref{1.3a}.
       The following two statements are equivalent:

       1) $L$  and $P$ generate $A_1$;

       2) $L$  and $P$ are bispectral.
       \epr

    \proof Let $L$ and $P$ be bispectral. According
    to \thref{1.2} $L$ is in the orbit of some
    $Q(\p)$, i.e. $L=\phi(Q(\p))$. Put $M = \phi^{-1}(P)$. Then the pair
    $(Q(\p),M)$ also satisfies \eqref{1.3a}. Obviously $M$ has at
    least one term depending on $x$. This automatically give that
    $Q$ is a polynomial of degree one, i.e. $Q=a\p +b, \quad a \neq 0$.
    Hence $M$ has the form $M= a^{-1}x +R(\p)$ with
    some polynomial $R$.  This shows that the pair $(Q(\p)),M$
    generate $A_1$. The same is true for their images $L,P$ under the
    automorphism $\phi^{-1}$, thus proving $2)\rightarrow 1)$.

    The opposite is obvious. Really. Let
    $L$ and $P$ generate $A_1$. Then they are strictly nilpotent,
    hence bispectral.



\renewcommand{\em}{\textrm}
\begin{small}
\renewcommand{\refname}{ {\flushleft\normalsize\bf{References}} }
    
\end{small}
\end{document}